%% file: ieee.tex
\documentclass[twocolumn,twosides,letterpaper,10pt]{IEEEtran}
\usepackage{cite,url,color}
\usepackage{graphicx} 
\graphicspath{{./pdf/}{./jpeg/}{./figs/}{./eps/}}
\DeclareGraphicsExtensions{.pdf,.jpeg,.png,.eps}
\usepackage[cmex10]{amsmath}
\usepackage{amssymb,bm}
\usepackage{algpseudocode} 
\newcommand{\algrule}[1][.5pt]{\par\vskip.25\baselineskip\hrule height #1\par\vskip.25\baselineskip}

\usepackage[caption=false,font=footnotesize]{subfig}
\usepackage[amsmath,thmmarks,hyperref]{ntheorem}

\usepackage{multirow}
\newcommand{\ve}[1]{\mathbf{{#1}}}
\newcommand{\abs}[1]{\ensuremath{\vert #1\vert}}
\newcommand{\norm}[1]{\ensuremath{\Vert #1\Vert}}

\begin{document}
%
\title{Energy Efficient Telemonitoring of Physiological Signals via Compressed Sensing: A Fast Algorithm and Power Consumption Evaluation}
%
%
%

\author{Benyuan~Liu$^{1,*}$,~Zhilin~Zhang$^{2,*}$,~Gary~Xu$^2$,~Hongqi~Fan$^1$,~Qiang~Fu$^1$%
\thanks{$^1$Benyuan Liu, Hongqi Fan and Qiang Fu are with The Science and Technology on Automatic Target Recognition Laboratory, National University of Defense Technology. Changsha, Hunan, P. R. China, 410074. E-mail: liubenyuan@gmail.com}%
\thanks{$^2$Zhilin Zhang and Gary Xu are with Samsung Research America -- Dallas. Richardson, TX 75082, USA. E-mail: zhilinzhang@ieee.org}%
\thanks{B. Liu and H. Fan were supported in part by the National Natural Science Foundation of China under Grant 61101186.}%
\thanks{Asterisk indicates corresponding author.}%
\thanks{Manuscript received \today{}.}}

\maketitle

\input{abstract.tex}

\ifCLASSOPTIONpeerreview
\fi
%
\IEEEpeerreviewmaketitle

\input{section01.tex  } 
\input{section02.tex  } 
\input{section03.tex  } 
\input{section04.tex  } 
\input{section05.tex  } 
\input{section06.tex  } 
\input{section07.tex  } 


%



\ifCLASSOPTIONcaptionsoff
  \newpage
\fi



\bibliographystyle{IEEEtran}
\bibliography{bsbl}
%



%






\end{document}

%% file: abstract.tex
%
\begin{abstract}
    Wireless telemonitoring of physiological signals is an important topic in eHealth. In order to reduce on-chip energy consumption and extend sensor life, recorded signals are usually compressed before  transmission. In this paper, we adopt compressed sensing (CS) as a low-power compression framework, and propose a fast block sparse Bayesian learning (BSBL) algorithm to reconstruct original signals. Experiments on real-world fetal ECG signals and epilepsy EEG signals showed that the proposed algorithm has good balance between speed and data reconstruction fidelity when compared to state-of-the-art CS algorithms. Further, we implemented the CS-based compression procedure and a low-power compression procedure based on a wavelet transform in Filed Programmable Gate Array (FPGA), showing that the CS-based compression can largely save energy and other on-chip computing resources.
\end{abstract}

%% file: section01.tex
\section{Introduction}
Monitoring physiological signals via wireless sensor networks is an important topic in wireless healthcare.
One major challenge of wireless telemonitoring is the conflict between huge amount of data collected and limited battery life of portable devices \cite{ZhangAsilomar2013,milenkovic2006wireless,duarte2012signal}. Data need to be compressed\cite{Higgins2013,duarte2012signal} before transmission. Most physiological signals are redundant, which means that they can be effectively compressed\cite{duarte2012signal} using transform encoders such as Discrete Wavelet Transform (DWT) based methods\cite{skodras2001jpeg}. However, these methods consist of sophisticated matrix-vector multiplication, sorting and arithmetic encoding which subsequently drain the battery.

\begin{figure}[!ht]
\centering
\includegraphics[width=3.4in]{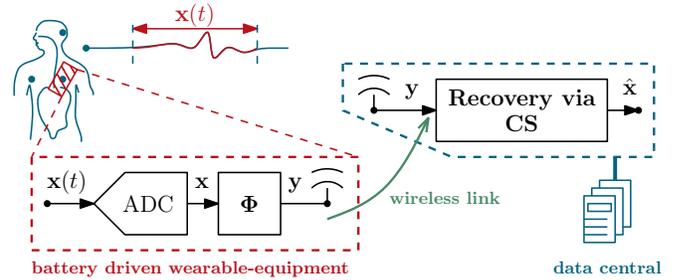}
\caption{The diagram of a Compressed Sensing (CS) based wireless telemonitoring system. }
\label{fig:diagram}
\end{figure}

Compressed Sensing (CS) \cite{Candes2008a}, can recover a signal with less measurements given that the signal is sparse or can be sparse represented in some transformed domains. CS-based wireless telemonitoring technology\cite{dixon2012compressed,mamaghanian2011compressed,Abdulghani2010,Zhang_TBME2012a,Zhang_TBME2012b,fauvel2014energy} can thus be viewed as a lossy compression method. The block diagram of a typical CS-based wireless telemonitoring is shown in Fig. \ref{fig:diagram}.
Physiological signals are firstly digitalized (Nyquist Sampling) via an Analog to Digital Converter (ADC). Those digitalized samples are compressed by a simple matrix-vector multiplication and the results are transmitted via wireless networks. At the data central, a CS algorithm is used to recover original signals from the compressed measurements.

\subsection{Overview of the Compressed Sensing}
The basic goal of  CS aims to solve the following underdetermined problem,
\begin{equation}\label{eq:bp}
\min \norm{\ve{x}}_1\quad s.t.\quad \ve{y} = \bm{\Phi}\ve{x},
\end{equation}
where $\ve{x}$ is the samples, $\bm{\Phi}$ is the sensing matrix whose row number is smaller than column number,  and $\ve{y}$ is the compressed measurements. $\norm{\ve{x}}_1$ is the $\ell_1$ norm penalty of $\mathbf{x}$, which prompts its sparsity.

In practice, physiological signals are not sparse  in the time domain, therefore one often resorts to a transformed domain such that $\mathbf{x}$ can be expressed as $\mathbf{x}=\mathbf{D}\bm{\theta}$ where $\ve{D}$ is a dictionary matrix such that the representation coefficients $\bm{\theta}$ are much sparser than $\mathbf{x}$. The problem in (\ref{eq:bp}) then becomes
\begin{equation}
\min \norm{\bm{\theta}}_1\quad s.t.\quad \ve{y} = (\bm{\Phi}\ve{D})\bm{\theta},
\end{equation}
The signal can be reconstructed afterwards using $\hat{\ve{x}}=\ve{D}\hat{\bm{\theta}}$ with the recovered coefficients $\hat{\bm{\theta}}$. Most CS-based telemonitoring systems\cite{mamaghanian2011compressed,Abdulghani2010} are build upon this model.

Recent advance in CS algorithms is to incorporate physical information \cite{Zhang2012a,polania2011compressed,polania2012exploiting} into the optimization procedure with the goal to achieve better reconstruction performance. One structure widely used is the block/group sparse structure\cite{Zhang2012a,Yuan2006,Baraniuk2010,Eldar2010}, which refers to the case when nonzero entries of a signal cluster around some locations. Moreover, noticing intra-block correlation widely exists in real-world signals, Zhang and Rao \cite{Zhang2012a,Zhang2011} proposed the block sparse Bayesian learning (BSBL) framework. It showed superior ability to recover block sparse signals or even non-sparse raw physiological signals such as fetal ECG \cite{Zhang_TBME2012b} and EEG signals \cite{Zhang_TBME2012a}.

\subsection{Summary of Contributions}
BSBL algorithms\cite{Zhang2012a} showed impressive recovery performance on physiological signals such as ECG and EEG. However, these algorithms derived so far are not fast and may limit their applications. The first contribution of our work is a fast implementation\footnote{The preliminary work of the developed algorithm is available in \url{http://arxiv.org/abs/1211.4909}.} of the BSBL framework using the Fast Marginalized (FM) likelihood maximization method\cite{Tipping2003}. Experiments conducted on real-life physiological signals showed that the proposed algorithm had similar recovery quality as BSBL algorithms, but was much faster.

Power consumption is a major concern in wireless telemonitoring systems. Traditionally, the power consumption was evaluated on a low-power Microcontroller (MCU)\cite{mamaghanian2011compressed}. However, MCU does not support fully parallel implementation and the power estimate is  affected by the coding style. In this work, we analyzed the power consumption on Field Programmable Gate Array (FPGA). In FPGA, we can implement the compressor in parallel and control the overall activities. Only the logic cells related to the compression core are implemented and the rest are holding reset. Therefore the power estimate is more accurate. In the experiment, the CS-based compressor was compared to a low-power DWT-based compressor in terms of compression latency, the number of utilized on-chip resources and power consumption. We proved that the CS-based architecture was more suitable for low-power physiological telemonitoring applications.

\subsection{Outline and Notations}
The rest of the paper is organized as follows. Section 2 presents the fast marginalized implementation of the BSBL algorithm and Section 3 provides the simulation setup and evaluation metrics. In Section 4 and Section 5, we conduct experiments on Fetal ECG (FECG) and EEG signals. The extracted FECGs and the Epileptic seizure classification results are used to evaluate the performance of CS. FPGA implementations and power consumption of the CS-based and the DWT-based compression methods are given in Section 6. Conclusion is drawn in the last section.

Throughout the paper, {\bf Bold} letters are reserved for vectors $\ve{x}$ and matrices $\ve{X}$. $\mathrm{Tr}(\cdot)$ computes the trace of a matrix and $\mathrm{diag}(\ve{A})$ extracts the diagonal vector of the matrix $\ve{A}$. $(\cdot)^T$ is the transpose operator. $\mathcal{N}(\ve{x};\bm{\mu},\bm{\Sigma})$ denotes a multivariate Gaussian distribution with mean $\bm{\mu}$ and variance $\bm{\Sigma}$.

%% file: section02.tex
\section{The Fast Implementation of The BSBL Framework}

\subsection{Overview of the BSBL Framework\cite{Zhang2012a}}
A block sparse signal $\ve{x}$ has the following structure,
\begin{equation}
\ve{x} = [\underbrace{x_1, \cdots, x_{d_1}}_{\ve{x}_1^T}, \cdots,
\underbrace{x_1, \cdots, x_{d_g}}_{\ve{x}_g^T}]^T,
\end{equation}
which means $\ve{x}$ has $g$ blocks, and only a few blocks are nonzero. Here $d_i$ is the block size for the $i$th block. The BSBL algorithms\cite{Zhang2012a} exploit the block structure and the intra-block correlation by modeling the signal block $\ve{x}_i$ using the parameterized Gaussian distribution:
\begin{equation}
p(\ve{x}_i;{\gamma_i},\ve{B}_i) =
\mathcal{N}(\ve{x}_i;\mathbf{0},{\gamma_i}\ve{B}_i). \label{eq:x_prior}
\end{equation}
with unknown deterministic parameters $\gamma_i$ and $\ve{B}_i$. $\gamma_i$ is a
nonnegative parameter controlling the block-sparsity of $\ve{x}$ and $\ve{B}_i$
is a positive definite matrix modeling the covariance structure of $\ve{x}_i$.
We assume that the blocks are mutually independent. Henceforth,
\begin{equation}\label{eq:x_model}
p(\ve{x};\{\gamma_i,\ve{B}_i\}_i) = \mathcal{N}(\ve{x};\ve{0},\bm{\Gamma}),
\end{equation}
where $\bm{\Gamma}$ denotes a block diagonal matrix with the $i$th principal block given by $\gamma_i\ve{B}_i$.

The measurement noise is assumed to be independent and Gaussian with zero mean and unknown variance $\beta^{-1}$. Thus the measurement model is
\begin{equation}
p(\ve{y}|\ve{x};\beta) = \mathcal{N}(\ve{y};\bm{\Phi}\ve{x},\beta^{-1}\ve{I}).
\label{eq:y_prior}
\end{equation}

Given the signal model \eqref{eq:x_model} and the measurement model \eqref{eq:y_prior},
the posterior $p(\ve{x}|\ve{y}; \{\gamma_i,\ve{B}_i\}_i,\beta)$ and the likelihood $p(\ve{y}|\{\gamma_i,\ve{B}_i\}_i,\beta)$ can be derived as follows,
\begin{align}
	p(\ve{x}|\ve{y}; \{\gamma_i,\ve{B}_i\}_i,\beta) &= \mathcal{N}(\ve{x};\bm{\mu},\bm{\Sigma}), \\
	p(\ve{y}|\{\gamma_i,\ve{B}_i\}_i,\beta) &= \mathcal{N}(\ve{y};\ve{0},\ve{C})
\end{align}
where
$\bm{\Sigma} \triangleq (\bm{\Gamma}^{-1} + \bm{\Phi}^T\beta\bm{\Phi})^{-1}$,
$\bm{\mu} \triangleq \bm{\Sigma}\bm{\Phi}^T\beta\ve{y}$ and
$\ve{C} \triangleq \beta^{-1}\ve{I} + \bm{\Phi}\bm{\Gamma}\bm{\Phi}^T$.
To estimate the parameters $\{\gamma_i,\mathbf{B}_i\}_i$ and $\beta$, the following cost function is used, which
is derived according to the Type II maximum likelihood \cite{Zhang2012a}:
\begin{align}
\mathcal{L}(\{\gamma_i,\mathbf{B}_i\}_i,\beta)
 &= -2\log p(\ve{y}|\{\gamma_i,\ve{B}_i\}_i,\beta) \\
 &\propto \log\abs{\ve{C}} + \ve{y}^T\ve{C}^{-1}\ve{y}, \label{eq:log-likelihood}
\end{align}

Once all the parameters are estimated, the MAP estimate of the signal $\mathbf{x}$ can be directly
obtained from the mean of the posterior, i.e., $\ve{x} = \bm{\mu}$.

\subsection{The Fast Implementation of BSBL}
There are several methods to minimize the cost function as shown in \cite{Zhang2012a}. In the following we consider to use the fast marginalized (FM) likelihood maximization method \cite{Tipping2003}.

The cost function \eqref{eq:log-likelihood} can be optimized in a block way. We denote by $\bm{\Phi}_i$ the $i$th block basis in $\bm{\Phi}$ with the column indexes  corresponding to the $i$th block of the signal $\ve{x}$. Then $\ve{C}$ can be rewritten as:
\begin{align}
\ve{C} &= \beta^{-1}\ve{I} + \sum_{m\neq i}
\bm{\Phi}_m\gamma_m\ve{B}_m\bm{\Phi}_m^T+
\bm{\Phi}_i\gamma_i\ve{B}_i\bm{\Phi}_i^T, \\
 &= \ve{C}_{-i} + \bm{\Phi}_i\gamma_i\ve{B}_i\bm{\Phi}_i^T, \label{eq:c}
\end{align}
where $\ve{C}_{-i} \triangleq \beta^{-1}\ve{I} + \sum_{m\neq i}
\bm{\Phi}_m\gamma_m\ve{B}_m\bm{\Phi}_m^T$.

Using the Woodbury Identity, $\abs{\ve{C}} = \abs{\ve{A}_i}\abs{\ve{C}_{-i}}\abs{\ve{A}_i^{-1} +
\ve{s}_i}$, $\ve{C}^{-1} = \ve{C}_{-i}^{-1} - \ve{C}_{-i}^{-1}\bm{\Phi}_i(\ve{A}_i^{-1} +
\ve{s}_i)^{-1}\bm{\Phi}_i^T\ve{C}_{-i}^{-1}$, where $\ve{A}_i\triangleq\gamma_i\ve{B}_i$,
$\ve{s}_i\triangleq \bm{\Phi}_i^T\ve{C}_{-i}^{-1}\bm{\Phi}_i$ and
$\ve{q}_i\triangleq\bm{\Phi}_i^T\ve{C}_{-i}^{-1}\ve{y}$,
Equation \eqref{eq:log-likelihood} can be rewritten as:
\begin{align}
\mathcal{L} =& \log\abs{\ve{C}_{-i}} + \ve{y}^T\ve{C}_{-i}^{-1}\ve{y} \nonumber
\\
 &+ \log\abs{\ve{I}_{d_i} + \ve{A}_i\ve{s}_i} - \ve{q}_i^T(\ve{A}_i^{-1} +
\ve{s}_i)^{-1}\ve{q}_i, \\
 =& \mathcal{L}(-i) + \mathcal{L}(i),
\end{align}
where $\mathcal{L}(-i) \triangleq \log\abs{\ve{C}_{-i}} +
\ve{y}^T\ve{C}_{-i}^{-1}\ve{y}$, and
\begin{equation}\label{eq:lgi}
\mathcal{L}(i) \triangleq
\log\abs{\ve{I}_{d_i} + \ve{A}_i\ve{s}_i} -
\ve{q}_i^T(\ve{A}_i^{-1} + \ve{s}_i)^{-1}\ve{q}_i,
\end{equation}
which only depends on $\mathbf{A}_i$.

Setting $\frac{\partial \mathcal{L}(i)}{\partial \ve{A}_i}=\ve{0}$, we have the updating rule
\begin{equation}\label{eq:a_0}
\ve{A}_i = \ve{s}_i^{-1}(\ve{q}_i\ve{q}_i^T - \ve{s}_i)\ve{s}_i^{-1}.
\end{equation}
In our model we further parameterize $\mathbf{A}_i$ as $\mathbf{A}_i = \gamma_i \mathbf{B}_i$ in order to impose some constraint (i.e. regularization) on the covariance structure (see the next subsection). However, there is ambiguity between $\gamma_i$ and $\mathbf{B}_i$. To solve this, we define $\gamma_i$ as the norm of $\mathbf{A}_i$, namely,
\begin{equation}\label{eq:gamma_0}
	\gamma_i = \norm{\ve{A}_i}_\mathcal{F},
\end{equation}
and define $\mathbf{B}_i$ as a matrix of unit norm which models the covariance structure of the $i$-th block $\mathbf{A}_i$, namely,
\begin{equation}\label{eq:b_0}
	\ve{B}_i = \frac{\ve{A}_i}{\gamma_i}.
\end{equation}
One can see this parameterization is a natural extension of the basic SBL model proposed by Tipping \cite{Tipping2001}.

\subsection{Imposing the Structural Regularization on $\mathbf{B}_i$}
\label{subsection_B}

As noted in \cite{Zhang2012a},  regularization to $\mathbf{B}_i$ is required due to limited data. A good regularization can largely reduce the probability of local convergence. Although theories on regularization strategies are lacking, some empirical methods \cite{Zhang2011,Zhang2012a} were presented.

In this paper we focus on the following two correlation models. One is the simple (SIM) model,
\begin{equation}
	\ve{B}_i=\ve{I}(\forall i).
\end{equation}
When the developed algorithm uses this regularization, we denote it by {\bf BSBL-FM(0)}. This model assumes that entries in each block are not correlated.

Another is to model the entries in each block as an Autoregressive (AR) process\cite{Zhang2011,Zhang2012a} of order $1$ with the coefficient $r_i$. The correlation level of the intra-block correlation is reflected by the value of $r_i$ which can be empirically calculated\cite{Zhang2012a} from the estimated $\ve{B}_i$ in (\ref{eq:b_0}). In many real-world applications, the intra-block correlation in each block of a signal tends to be positive and high together. Thus, we further constrain that all the intra-block correlation values of blocks have the same AR coefficient $r$ \cite{Zhang2012a},
\begin{align}
r &= \frac{1}{g}\sum_{i=1}^g r_i.  \label{equ:r}
\end{align}
Then $\mathbf{B}_i$ is reconstructed as a symmetric Toeplitz matrix with the first row given by
$[1,r,\cdots,r^{d_i-1}]$.
Our algorithm using this regularization is denoted by {\bf BSBL-FM(1)}.


\subsection{Remarks on $\beta$}
The parameter $\beta^{-1}$ is the noise variance in our model. It can be
estimated by solving $\frac{\partial \mathcal{L}}{\partial \beta}$ from \eqref{eq:log-likelihood},
\begin{equation}
\beta = \frac{M}{\mathrm{Tr}[\bm{\Sigma}\bm{\Phi}^T\bm{\Phi}] +
\norm{\ve{y}-\bm{\Phi}\bm{\mu}}_2^2}.
\end{equation}
However, the resulting updating rule is generally not robust and thus
requires some regularization \cite{Zhang2012a,Zhang2011}. Alternatively, one can treat it as a regularizer and assign suitable values to it via cross-validation. In our experiments we set 
$\beta^{-1}=10^{-6}$ in noiseless simulations and
$\beta^{-1}=0.01\norm{\mathbf{y}}_2^2$ in noisy scenarios, which showed good performance.

\subsection{The Fast Marginalized block SBL Algorithm}
The Fast Marginalized block SBL algorithm
(BSBL-FM) is given in Fig. \ref{algo:bsbl-fm}.
\begin{figure}[h!]
\centering
\begin{algorithmic}[1]
    \algrule
\Procedure{BSBL-FM}{$\ve{y}$,$\bm{\Phi}$,$\eta$}
\State Outputs: $\ve{x},\bm{\Sigma}$
\State Initialize: $\beta^{-1} = 0.01\|\ve{y}\|_2^2$ (noisy cases) or $\beta^{-1}=10^{-6}$ (noiseless cases),  $\ve{B}_i=\ve{I}$,  $\ve{s}_i=\beta\bm{\Phi}_i^T\bm{\Phi}_i, \ve{q}_i=\beta\bm{\Phi}_i^T\ve{y}, \forall i$
\While{not converged}
\State Calculate $\ve{A}'_i= \ve{s}_i^{-1}(\ve{q}_i\ve{q}_i^T - \ve{s}_i)\ve{s}_i^{-1}, \forall i$
\State Calculate $\gamma_i = \norm{\ve{A}'_i}_\mathcal{F}$
\State Imposing Structural Regularization on $\ve{B}_i$
\State Re-build $\ve{A}^*_i = \gamma_i\ve{B}^*_i$
\State Calculate $\Delta \mathcal{L}(i) = \mathcal{L}(\ve{A}^*_i) - \mathcal{L}(\ve{A}_i), \forall i$
\State Select the $\hat{i}$th block s.t. $\Delta\mathcal{L}(\hat{i})=\min\{\Delta\mathcal{L}(i)\}$
\State Re-calculate $\bm{\mu},\bm{\Sigma},\{\ve{s}_i\},\{\ve{q}_i\}$
\EndWhile
\EndProcedure
    \algrule
\end{algorithmic}
\caption{The Proposed BSBL-FM Algorithm.}
\label{algo:bsbl-fm}
\end{figure}
Within each iteration, it only updates the block signal that attributes to the deepest descent of $\mathcal{L}(i)$. The detailed procedures on re-calculation of $\bm{\mu},\bm{\Sigma},\{\ve{s}_i\},\{\ve{q}_i\}$ are similar to \cite{Tipping2003}. The algorithm terminates when the maximum change of the cost function is smaller than a threshold $\eta$. In our experiments, we set $\eta=1\mathrm{e}^{-5}$.



%% file: section03.tex
\section{System Evaluation Settings}

\subsection{The Experiments Setup}
In our experiments, physiological signals were divided into packets. Each packet consisted of digitalized samples collected within a time window $t_p$ where $t_p=Nt_s$, $N$ was the number of samples within a packet and $t_s$ was the sampling interval. The packet was compressed by multiplying a sparse binary matrix $\bm{\Phi}$ of size $M\times N$ with $k$ non-zero entries of each column. Note that the sparse binary sensing matrix\cite{mamaghanian2011compressed,Zhang_TBME2012a,Zhang_TBME2012b} has been widely used in CS-based telemonitoring for its efficiency in storage and matrix-vector multiplication. The compression ratio ($\mathrm{CR}$) was defined as
\begin{equation}
	\mathrm{CR} = \frac{N-M}{N}.
\end{equation}

\subsection{The Compared CS Algorithms}
We compared the proposed algorithm with state-of-the-art CS algorithms. The algorithms and their features are listed in Table \ref{tab:cs}.
\begin{table}[!ht]
\renewcommand{\arraystretch}{1.3}
\centering
\caption{The CS recovery algorithms used in this paper.}%
\label{tab:cs}
\begin{tabular}{lcc}
\hline\hline
{\bf CS Algorithm} & {\bf Objective Function} & {\bf Features} \\
\hline
BP\cite{Kim2007}            & $\ell_1$ minimization & \\
Group BP\cite{Yuan2006}     & group $\ell_1$ minimization & block \\
\hline
BSBL-BO\cite{Zhang2012a}    & BSBL based & block + correlation \\
BSBL-FM(0)         & fast BSBL based & block \\
BSBL-FM(1)         & fast BSBL based & block + correlation \\
\hline\hline
\end{tabular}
\end{table}
Throughout the experiments, the number of samples within a packet was fixed to $N=512$. Default parameters for BP were used. For Group BP, we tuned the parameters `optTol' to $\mathrm{10}^{-5}$ and `iterations' to $200$ for optimum performance. $\beta^{-1}=\mathrm{10}^{-6}$ was selected for BSBL based algorithms. The same block partition (block size equals to $32$) was used for all block based recovery algorithms.

\subsection{The Performance Indexes}
Throughout the experiments, we used two performance indexes. One was the Percentage root-mean squared distortion (PRD), defined as
\begin{equation}
	\mathrm{PRD} = \frac{\norm{\ve{x} - \ve{\hat{x}}}_2}{\norm{\ve{x}}_2} \cdot 100,
\end{equation}
where $\ve{\hat{x}}$ was the reconstructed signal of the true signal $\ve{x}$.
The lower the PRD, the better the recovery performance.
Another was the CPU time, which was calculated on a computer with 2.9GHz CPU and 16G RAM.

Two experiments were carried out. One was Fetal ECG (FECG) telemonitoring. Similar as in \cite{Zhang_TBME2012a,Zhang_TBME2012b}, we compared the Independent Component Analysis (ICA) decomposition \cite{hyvarinen1999fast} of original FECG recordings to the ICA decomposition of recovered FECG recordings. %

Another experiment was EEG telemonitoring for epileptic patients. We evaluated the distortion of CS by comparing the seizure classification results. The metric, called Area Under the receiver operation Curve (AUC) \cite{Higgins2013}, was calculated to evaluate the classification performance. AUC denotes the area below the plot of sensitivity (true positive rate) versus specificity (true negative rate).

%% file: section04.tex

\section{Application to Telemonitoring of Fetal Electrocardiogram}

The FECG dataset used in the experiment was the same as in Section III.B of \cite{Zhang_TBME2012b} \footnote{Available on-line: https://sites.google.com/site/researchbyzhang/bsbl}. The FECG dataset consisted of eight abdominal (channels) recordings sampled at $250$Hz and each recording contained $10240$ data points.

\begin{figure}[!ht]
\centering
\includegraphics[width=3.2in]{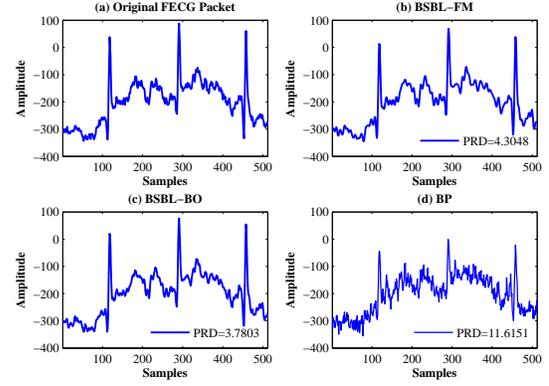}
\caption{(a) A raw FECG packet. (b), (c) and (d) are the recovered signals by BSBL-FM, BSBL-BO and BP, respectively. The compression ratio $\mathrm{CR}=0.60$ and $\mathbf{D}$ was a discrete cosine transform dictionary matrix. Note that the significant ``spikes'' are the QRS complex waves of the maternal ECG. The desired fetal ECG is buried in the baseline noise and the maternal ECG.}
\label{fig:fecg_sample}
\end{figure}

An example on a raw FECG packet reconstructed by different CS algorithms is shown in Fig. \ref{fig:fecg_sample}.
In this example, a FECG packet of $N=512$ samples was compressed with $\mathrm{CR}=0.60$. The recovered signals as well as the distortions using different CS algorithms are shown in Fig. \ref{fig:fecg_sample}.

In the following we analyzed various factors that affected the distortion of CS, including the choice of the dictionary matrix $\ve{D}$ and the compression ratio $\mathrm{CR}$. The ICA decompositions on the recovered recordings were used to verify the design results.

\subsection{The Dictionary Matrix $\ve{D}$}
\begin{figure}[!ht]
\centering
\includegraphics[width=3.2in]{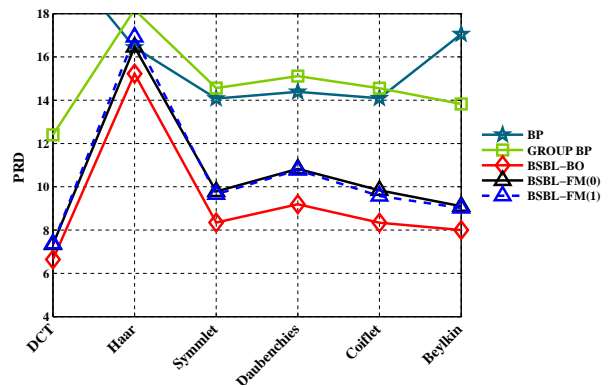}
\caption{The performance comparison of CS algorithms in recovering the whole Fetal ECG recordings using different dictionary matrices.}%
\label{fig:fecg_D}
\end{figure}

The dictionary matrix $\ve{D}$ was used to sparsely represent the physiological signal $\ve{x}$ in a transformed domain. In this experiment, we considered six orthogonal dictionary matrices. One was the Discrete Cosine Transform (DCT) matrix, which has been widely used in biological signal processing\cite{Zhang_TBME2012a,Zhang_TBME2012b}. The next five were wavelet transform matrices\footnote{The wavelet transform matrix $D$ was generated using the {\it wavemat} function of the {\it wavelab} toolbox, available at http://www-stat.stanford.edu/\~{}wavelab/}, namely, the Haar wavelet, the Symmlet wavelet (the number of vanishing moments was set to $6$), the Daubenchies wavelet (the number of vanishing moments was set to $8$), the Coiflet wavelet and the Beyklin wavelet. In this experiment, $N=512$, $k=2$ and $\mathrm{CR}=0.60$. The results are shown in Fig. \ref{fig:fecg_D}.

From Fig. \ref{fig:fecg_D} we found that the algorithms derived from the BSBL framework had better performance than the other recovery algorithms. The best performance of the BSBL family was obtained when $\ve{D}$ was the DCT matrix.

\subsection{The Compression Ratio $\mathrm{CR}$}
\begin{figure}[!ht]
\centering
\includegraphics[width=3in]{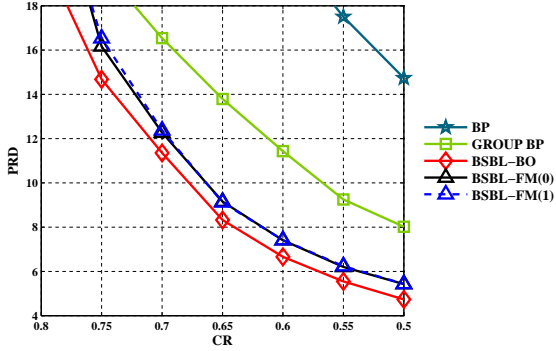}
\caption{The PRDs of different CS algorithms with varying compression ratios $\mathrm{CR}$.}%
\label{fig:fecg_k}
\end{figure}

\begin{figure}[!ht]
\centering
\includegraphics[width=3in]{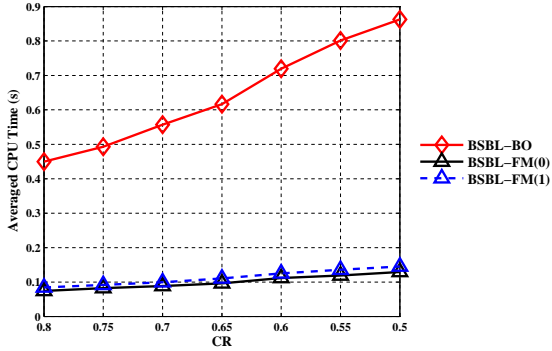}
\caption{The average CPU times of different BSBL algorithms with varying compression ratios $\mathrm{CR}$.}%
\label{fig:fecg_cpu}
\end{figure}

The data distortion from CS is also affected by the compression ratio $\mathrm{CR}$. As more measurements are obtained, a CS algorithm's recovery performance can be further improved. In this experiment we studied the recovery performance of all compared algorithms in terms of CR. The dictionary matrix $\ve{D}$ was a DCT matrix,  $N=512$, and $k=2$. The results are shown in Fig. \ref{fig:fecg_k} and Fig. \ref{fig:fecg_cpu}.

From Fig. \ref{fig:fecg_k} we found that the proposed algorithm showed similar performance as BSBL-BO, and all BSBL algorithms significantly outperformed other algorithms.

From Fig. \ref{fig:fecg_cpu}, we may conclude that our fast implementation was in average $\sim7$ times faster than BSBL-BO. We also noted that, in the DCT transformed domain, the coefficients $\bm{\theta}$ were almost not correlated. Therefore, the algorithms BSBL-FM(0) and BSBL-FM(1) yielded similar performances. However, by regularizing $\ve{B}_i=\ve{I}$, BSBL-FM(0) reduced computational load compared to BSBL-FM(1).

\subsection{The ICA Decomposition of Recovered FECG Recordings}
\begin{figure}[!ht]
\centering
\includegraphics[width=3.5in]{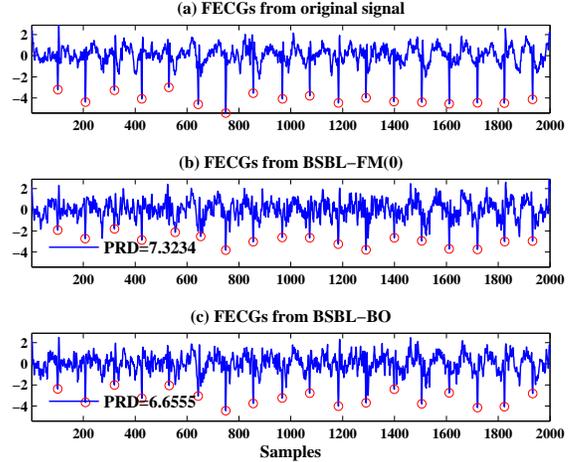}
\caption{Extracted clean FECGs using ICA (a) from the original raw FECG recordings, (b) from the recovered raw FECG recordings by BSBL-FM(0), and (c) from the recovered raw FECG recordings by BSBL-BO. Only the first $2000$ sampling points of the datasets are shown in the figure. The averaged PRDs in recovering the whole Fetal ECG recordings are shown in the corner of subfigure (b) and (c). The red circles denote the detected peaks of the FECGs using the MATLAB function \emph{findpeaks} where `MinPeakHeight' was set to $1.6$ and `MinPeakDistance' was set to $80$.}
\label{fig:fecg_ica}
\end{figure}

As illustrated in Fig. \ref{fig:fecg_sample}, the desired FECG was buried in baseline noise and maternal ECG. Thus, we used ICA to extract the clean FECG from the raw FECG recordings recovered by different CS algorithms. In this experiment, $\mathrm{CR}=0.60$, $k=2$, $N=512$ and $\ve{D}$ was a DCT dictionary matrix. The extracted FECGs of different CS algorithms are shown in Fig. \ref{fig:fecg_ica} and the average PRDs and the CPU time of these algorithms are shown in Table \ref{tab:fecg_ica}. We can see that although BSBL-FM(0) had slightly poorer recovery performance than BSBL-BO, it was much faster. Furthermore, as shown in Fig.\ref{fig:fecg_ica}, the extracted clean FECG using ICA from the recovered recordings by BSBL-FM(0) was almost the same as the extracted one from the original recordings.

\begin{table}[!htb]
	\renewcommand{\arraystretch}{1.3}
	\centering
	\caption{The average PRD and CPU time in recovering the FECG recordings by all algorithms. }
	\label{tab:fecg_ica}
	\begin{tabular}{ccccc}
		\hline\hline
		& BP & Group BP & BSBL-BO & \textbf{BSBL-FM(0)} \\
		\hline
		\textbf{PRD} &  20.98 & 11.43 & {\bf 6.65} & {\bf 7.32} \\
		\textbf{CPU Time (s)} & 0.074 & 0.126 & 0.620 & {\bf 0.122} \\
		\hline\hline
	\end{tabular}
\end{table}

%% file: section05.tex
%
\section{Applications to EEG Telemonitoring for the Epileptic Patients}
Epilepsy is a common chronic neurological disorder. About one out of every three patients with epilepsy continue to experience frequent seizures\cite{shoeb2009application}. Using EEG signals as the proxy to identify and detect epilepsy seizures has been widely studied\cite{song2010new,yuan2011epileptic,shoeb2009application,shoeb2010application}. A low-energy designed telemonitoring framework is valuable for such application as it can provide all-day long continuous monitoring, which makes epilepsy more likely to be detected and suitably treated.

\subsection{Description of the EEG datasets}
The EEG dataset used in this experiment was the same as \cite{goldberger2000physiobank,shoeb2009application}. It consisted of EEG recordings with intractable seizures from $22$ subjects. For each subject, there were hours of recordings with clinician annotated seizure segments. Such annotations were used to evaluate the epileptic classification results.
The EEG dataset was sampled at $256$Hz with $16$-bit resolution. Our objective here is to evaluate the distortion of lossy compression via CS.

\subsection{Experiments Setup}
In the experiment, EEG signals were divided into packets. Each packet was 2s long which contained $N=512$ samples. The $i$th packet was denoted as $\{\ve{x}_i^1,\cdots,\ve{x}_i^C\}$ where $C$ was the number of electrodes and $\ve{x}_i^j$ was the $i$th packet from electrode $j$. A binary sensing matrix $\bm{\Phi}$ with two entries of 1s in each column (k=2) was generated. Each packet in $\{\ve{x}_i\}$ was compressed using this sensing matrix, resulting in the compressed measurements $\{\ve{y}_i\}$.

\begin{figure}[!ht]
	\centering
    \includegraphics[width=3in]{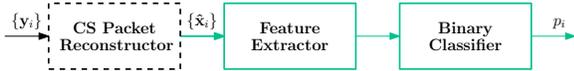}
	\caption{The epileptic seizure classifier block. The compressed EEG packets received at the data central are denoted by $\{\ve{y}_i\}$. The original EEG signals $\{\ve{\hat{x}}_i\}$ are reconstructed from $\{\ve{y}_i\}$ using CS solvers and piped through the feature extractor and the classifier. The classifier outputs a probability $p_i$ which denotes abnormality of the current packet.}
	\label{fig:epileptic_classifier}
\end{figure}

We used machine learning techniques to automatically detect the epileptic seizure packets. The diagram of the epileptic seizure classifier is shown in Fig. \ref{fig:epileptic_classifier}.
The features used in this paper were non-linear features\cite{yuan2011epileptic,song2010new}, namely Approximate Entropy\cite{acharya2012automated}, Sample Entropy \cite{aboy2007characterization}, Hurst Exponential \cite{acharya2005non}, and Scale Exponential \cite{leistedt2007characterization}. These features were fed into a Random Forest (RF) \cite{criminisi2011decision} classifier to classify the epileptic seizure segments.

\subsection{Epileptic Seizure Classification Results at Different CR Values}
\begin{figure}[!ht]
	\centering
	\includegraphics[width=3in]{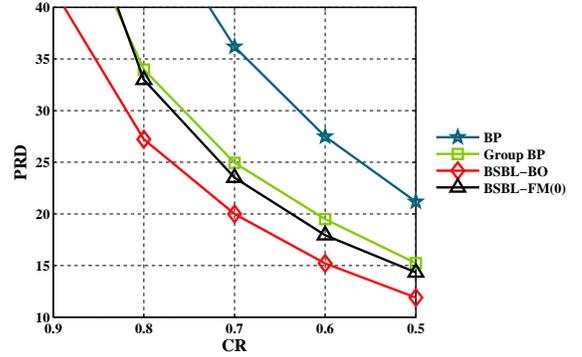}
    \caption{The PRD in recovering EEG signals with different CR values. With $\mathrm{CR}=0.80$, the averaged CPU time in recovering the whole EEG recordings were 0.071s for BP, 0.163s for Group-BP, 0.420s for BSBL-BO and \textbf{0.097s} for BSBL-FM(0).}
	\label{fig:eeg_prd}
\end{figure}

The distortion introduced by CS varies with different CS solvers and different CR values. In this experiment, we varied $\mathrm{CR}$ from $0.5$ to $0.9$ and calculated the average PRD of each algorithm. The result is shown in Fig. \ref{fig:eeg_prd}. Again we saw that BSBL-FM(0) had similar recovery performance as BSBL-BO but was much faster.

\begin{figure}[!ht]
	\centering
	\includegraphics[width=3in]{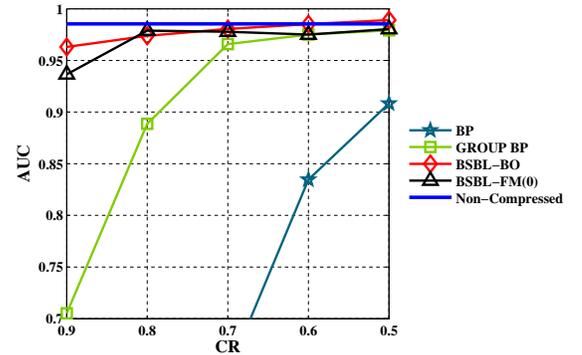}
    \caption{The classification performance index AUC on the recovered EEG signals by all CS algorithms at different CR values. For better comparison, the AUC calculated on the original EEG signals is also shown and denoted by `non-compressed'.}
	\label{fig:eeg_auc}
\end{figure}

The AUC of the epileptic seizure classifier on the recovered signals was also calculated. The classifier was trained on 80\% of the recovered dataset and tested on the rest dataset. Fig. \ref{fig:eeg_auc} shows the calculated AUCs when the classifier performed on the recovered signals by all algorithms at different CR values. The result showed that BSBL-FM(0) had similar recovery quality as BSBL-BO. All the BSBL based algorithms had comparable AUC metrics to the `non-compressed' with CR ranging from $0.50$ to $0.90$.

%% file: section06.tex
\section{Hardware Implementation and Evaluation}

\subsection{Background}
We implemented data compressors in FPGA and evaluated the power consumption. Unlike Microcontrollers (MCU), the FPGA supports fully parallel implementation and compact design. It uses only the logic cells related to the compressor while the rest are holding reset, therefore the power estimate is more accurate.

A typical design flow on FPGA involves mapping the high-level description language into the real-time logic cells in the chip. For Xilinx FPGAs, the basic cells are Flip-Flops (denoted as FFs), Look-Up-Tables (LUTs), Dual-port memories (denoted as RAMs) and DSP48 slices\footnote{The basic memory primitive of Xilinx Spartan 6 FPGA is RAM8BWER. A RAM8BWER (RAM for short) provides a total memory of 9Kb. The DSP48 slices are digital signal processing logic elements. A DSP48 slice can perform fixed-point arithmetic operations such as multiply-accumulator, multiply-adder, etc.}. In order to achieve low-power consumption, the design should utilize less resources, have low Flip-Flop toggling rates and be multiplierless.

\subsection{Implementation}
Two compression systems for telemonitoring physiological signals were implemented.
One was a traditional DWT-based and another was our CS-based architecture.

\subsubsection{DWT-based physiological signal compression}
\begin{figure}[!ht]
    \centering
    \includegraphics[width=2.8in]{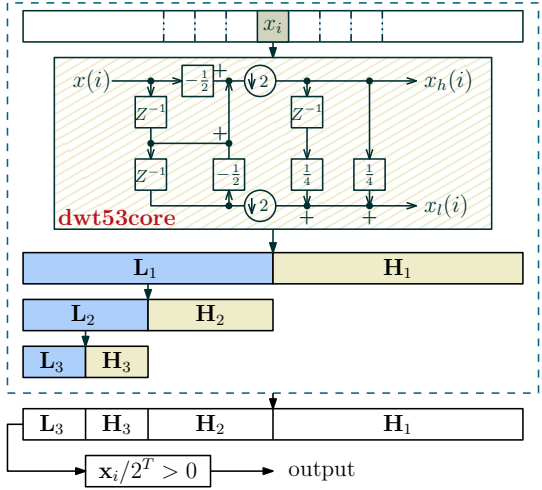}
    \caption{The implementation of CDF 5/3 wavelet compression core. `dwt53core' implements the lift-based filtering. $\ve{L}$ and $\ve{H}$ denote the storage for low-passed and high-passed coefficients respectively.}
    \label{fig:dwt_arch}
\end{figure}

A DWT based one dimensional (1D) signal compression core was implemented using the Cohen-Daubechies-Feauveau (CDF) 5/3 wavelet filter. This biorthogonal wavelet is multiplierless and it also supports lift-based filtering\cite{skodras2001jpeg}, as shown in Fig. \ref{fig:dwt_arch}. We also implemented multiple transformation stages to successively improve the resolution.
The high-pass coefficients $\ve{x}_h$ and the low-pass coefficients $\ve{x}_l$ are computed as
\begin{align}
x_h(i) &= x(2i) + \Big\lfloor -\frac{1}{2}\left[x(2i-1) + x(2i+1)\right] \Big\rfloor, \\
x_l(i) &= x(2i-1) + \Big\lfloor \frac{1}{4}\left[x_h(i-1) + x_h(i)\right] + \frac{1}{2} \Big\rfloor,
\end{align}
where $\lfloor\cdot\rfloor$ is the floor function. $x(2i)$, $x(2i+1)$ and $x(2i-1)$ are data samples or the low-passed DWT coefficients of a previous stage. The multiplication with $1/2$ and $1/4$ in the lifting process can be efficiently implemented in FPGA using shifting operations. Therefore, no multiplier is used.

The compression is achieved by passing the DWT coefficients through a threshold testing module. This module discards all  coefficients with encoding bits  less than $T$, i.e., $\abs{x_i/2^T} \leq 0$. Finally, both the coefficients and their corresponding locations are outputted.

\subsubsection{CS-based physiological signal compression}
\begin{figure}[!ht]
    \centering
    \includegraphics[width=2.8in]{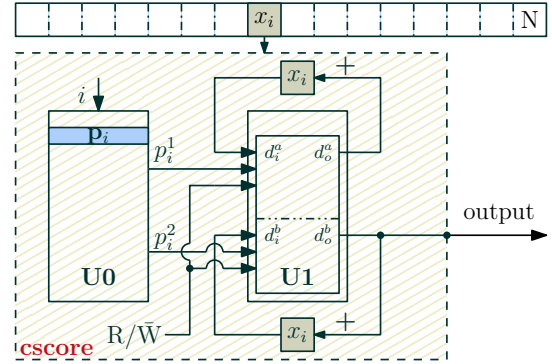}
    \caption{The implementation of the CS-based compressor ($k=2$). Block U0 is a Read-Only-Memory (ROM) block that stores the locations of 1s in the sensing matrix. Block U1 is a synchronous Dual-port memory that stores the compressed measurements $\ve{y}$. $d_i^{a,b}$ and $d_o^{a,b}$ denote the data inputs and outputs of U1. When a new sample $x_i$ arrived, the system reads out the contents in U1 where $\{p_i^1,p_i^2\}$ point to, accumulates by $x_i$ and write them back. U1 is cleared after being unloaded for the compression of the next packet.}
    \label{fig:cs_arch}
\end{figure}

The CS-based compression is expressed as $\ve{y}=\bm{\Phi}\ve{x}$, where $\bm{\Phi}$ is the sparse binary matrix with each column containing only two entries of 1s. Our implementation can compress the samples \emph{on-the-fly},
\begin{equation}
    \ve{y} = \bm{\Phi}\ve{x} = \sum_{i=1}^{N} \bm{\phi}_i x_i,
\end{equation}
where $\bm{\phi}_i$ is the $i$th column vector of the sensing matrix.  Let $\ve{y}^{(k)}$ be the compressed measurements after the $k$th sample $x_k$ has been collected, i.e., $\ve{y}^{(k)}\triangleq\sum_{i=1}^k \bm{\phi}_ix_i$. Starting with $\ve{y}^{(0)}=\ve{0}$, the compressed measurements are iteratively updated with each new sample $x_i$ available,
\begin{equation}\label{eq:cs_fpga}
	y_{p_i^1}^{(i)} = y_{p_i^1}^{(i-1)} + x_i, \quad y_{p_i^2}^{(i)} = y_{p_i^2}^{(i-1)} + x_i.
\end{equation}
where $\{p_i^1,p_i^2\}$ are the indexes of 1s in $\bm{\phi}_i$. \eqref{eq:cs_fpga} can be calculated in parallel with no multipliers. The implementation is shown in Fig. \ref{fig:cs_arch}.


\subsection{Evaluation}
In the experiment, the number of stages of the DWT-based compressor was $4$ and $T$ in the threshold testing module was $8$. The compression ratio $\mathrm{CR}$ of the CS-based compressor was $0.50$. We stored 20 packets of EEG recordings in a ROM and each data sample was quantized with 16-bit resolution. The FPGA sequentially read out the data and compressed them using the two compressors we had implemented. The on-chip signals were monitored using the chipscope logic analyzer via a debug cable. We then exported the compressed measurements captured by chipscope in the Value Change Dump (VCD) file format. These VCD files were used in MATLAB to assess the distortion of data recovery.

\subsubsection{On-chip Activities}
On-chip activities of a compressor are mainly reflected by the compression latency and the toggling rates. The compression latency, denoted by $t_l$, is the number of clock cycles a core takes to compress a packet. The toggling rates, denoted by $T_r$, represents how often the signals in FPGA toggle their values with respect to the system clock. For example, a signal toggles its value at every rising edge of the system clock has a toggling rates equal to $100$\%. The FPGA chip is dynamic within the toggling times and static (idle) otherwise. The compression latency and the toggling rates can be precisely calculated using the VCD files monitored by chipscope.

The CS-based implementation compresses the data on-the-fly, therefore the compression ends immediately after the last sample of a packet. In contrast, the DWT-based implementation operates in a batch mode
and it requires a whole packet to perform the multistage data transformation. The latency $t_l$ of a 4-stage DWT core can be roughly calculated as $t_l \approx (N + N/2 + N/4 + N/8)$ where $N$ represents the number of samples within a packet.

In our experiment, the compression latencies $t_l$ and the toggling rates $T_r$ monitored by chipscope are shown in Table \ref{tab:fpga}. The CS-based compressor had minimal compression latency and it consumed only 34.3\% of on-chip activities of the DWT-based compressor.

\subsubsection{Resource and Power Consumption}
The utilized on-chip resources are shown in Table \ref{tab:fpga}. The FFs and LUTs consumed by the CS-based compressor were only 32.7\% and 33.1\% of the DWT-based one. The CS-based compressor also had less RAM usage. This is because the DWT-based compressor used RAMs to store the transformed coefficients and mid-stage filtering results, while for the CS-based compressor, only the locations of 1s in the sensing matrix and the compressed results were stored.

We used Xilinx Power Analyzer (XPA) to estimate the dynamic power $P_d$ of the compressor. The dynamic power was associated with design complexity and switching events in the core. From the power estimates in Table \ref{tab:fpga}, we found that the dynamic power consumed by the CS-based compressor was $68.7$\% of the DWT-based one. Further, we denote by $P_1$ the energy consumption of FPGA in compressing one packet, which is calculated by
\begin{equation}
	P_1 = P_d\cdot t_p\cdot T_r.
\end{equation}
From the results in Table \ref{tab:fpga}, the energy consumption of the CS-based compressor was only $23.7$\% of the DWT-based one.

\begin{table}[!ht]
\renewcommand{\arraystretch}{1.3}
\centering
\caption{The hardware evaluation of the DWT-based compressor and the CS-based compressor. The compression latency $t_l$, toggling rates $T_r$, the number of utilized on-chip resources (flip-flops (FF), looking-up-tables (LUT), block memories (RAM)), dynamic power consumption $P_d$ and the energy consumption in compressing one packet $P_1$ were compared.}
\label{tab:fpga}
\begin{tabular}{c|cc|ccc|cc}
\hline\hline
& $t_l$ & $T_r$ & FFs & LUTs & RAMs & $P_d$ & $P_1$ \\
\hline
DWT & 974 & 0.297\% & 223 & 359  & 4     & 16mW  & 95.3uJ \\
CS  &{\bf 2}&{\bf 0.102\%}&{\bf 73}&{\bf 119}&{\bf 3}&{\bf 11mW}&{\bf 22.6uJ}\\
\hline\hline
\end{tabular}
\end{table}

%% file: section07.tex
\section{Conclusion}
This work proposed a fast block sparse Bayesian learning algorithm, called BSBL-FM, for compressed sensing of physiological signals. Experiments on fetal ECG data and epileptic EEG data showed that the algorithm had similar recovery performance as other BSBL algorithms, but was much faster.
Furthermore, the compression procedures of compressed sensing and a low-power wavelet compression algorithm were implemented in FPGA, and were compared in terms of power and energy consumption and other on-chip computing resources. The comparison results showed that  energy consumption and dynamic power consumption of the CS compression procedure were only 23.7\% and 68.7\% of those of the wavelet compression procedure, respectively. Besides, on-chip computing resources including flip-flops, looking-up-tables, and  memories utilized by the CS compression were also largely reduced.
Those results suggest that the proposed algorithm is very suitable for energy-efficient real-time ambulatory telemonitoring of physiological signals.